\documentclass[prl,twocolumn,nofootinbib,superscriptaddress]{revtex4-1}
\usepackage{graphicx}
\usepackage{dcolumn}
\usepackage{bm,bbm}
\usepackage{xcolor}
\usepackage{tcolorbox}
\usepackage{algorithm}
\usepackage{algpseudocode}
\usepackage{amsmath}
\usepackage{braket}
\usepackage{nameref}
\usepackage[toc,page]{appendix}

\usepackage[colorlinks,breaklinks,linkcolor={blue},citecolor={magenta},urlcolor={blue}]{hyperref}
\usepackage{enumerate}
\usepackage{array}
\usepackage{booktabs}
\usepackage{amsmath,amssymb,amstext,amsfonts} 
\usepackage{mathtools}
\usepackage{cancel}
\usepackage{siunitx}  
\usepackage{physics}  
\usepackage{amsthm}  

\makeatletter
\newcommand\org@hypertarget{}
\let\org@hypertarget\hypertarget
\renewcommand\hypertarget[2]{%
  \Hy@raisedlink{\org@hypertarget{#1}{}}#2%
  }
\makeatother
\begin{document}

\title{A genetic algorithm for the response of \\ twisted nematic liquid crystals to an applied field}

\author{Alicia  Sit}
\thanks{Current affiliation: National Research Council of Canada, 100 Sussex Drive, K1N 5A2, Ottawa, Ontario, Canada}
\affiliation{Nexus for Quantum Technologies, University of Ottawa, K1N 5N6, Ottawa, ON, Canada}

\author{Francesco Di Colandrea}
\email{francesco.dicolandrea@uottawa.ca}
\affiliation{Nexus for Quantum Technologies, University of Ottawa, K1N 5N6, Ottawa, ON, Canada}

\author{Alessio  D'Errico}
\affiliation{Nexus for Quantum Technologies, University of Ottawa, K1N 5N6, Ottawa, ON, Canada}

\author{Ebrahim Karimi}
\affiliation{Nexus for Quantum Technologies, University of Ottawa, K1N 5N6, Ottawa, ON, Canada}


\begin{abstract}
When an external field is applied across a liquid-crystal cell, the twist and tilt distributions cannot be calculated analytically and must be extracted numerically. In the standard approach, the Euler-Lagrange equations are derived from the minimization of the free energy of the system and then solved via finite-difference methods, often implemented in commercial software. These tools iterate from initial solutions that are compatible with the boundary conditions, providing limited to no flexibility for customization. Here, we present a genetic algorithm that outputs fast and accurate solutions to the integral form of the equations. In our approach, the evolutionary routine is sequentially applied at each position within the bulk of the cell, thus overcoming the necessity of assuming trial solutions. The predictions of our routine strongly support the experimental observations on different instances of spatially varying twisted nematic liquid-crystal cells, patterned with different topologies on the two alignment layers.
\end{abstract}

\date{\today}
\maketitle

\section*{Introduction}
The response of liquid crystals to an applied field, whether magnetic~\cite{leslie:70} or electric~\cite{deuling:74}, can be modeled within the elastic continuum theory~\cite{gennes:93}. This predicts that the molecules tend to align with the field direction. The equilibrium configuration can be found by imposing the minimization of the total free energy. At the equilibrium, the twist and tilt distributions of the liquid-crystal director thus satisfy a set of Euler-Lagrange equations~\cite{deuling:74,preist:89}. These are typically solved via finite-difference or finite-element methods~\cite{yang2014fundamentals,Moser:19}, mostly embedded in modeling software such as LC3D~\cite{Wang:04,Wang:05}, DIMOS~\cite{Peng:16,lee2017two} and COMSOL~\cite{guirado2022dynamic,gorkunov2022double,sova2023theoretical}. Starting from trial distributions that satisfy the boundary conditions, these routines converge to quasi-optimal solutions within a range of iterations.

A recent work explored the application of genetic algorithms to determine the director distribution in a few relevant cases, including hybrid and twisted nematic liquid-crystal cells~\cite{yang2020genetic}. 
Inspired by Darwin's evolutionary theory, the basic idea is to start from a population of random guesses (\textit{individuals}) and let the workflow of the genetic algorithm (GA) select the individuals that better approximate the optimal solution of the physical problem~\cite{holland:92}, formulated in terms of the minimization of a cost function (\textit{fitness}). The workflow is based on the application of an iterative scheme composed of three sequential operators: \textit{selection}, \textit{crossover}, and \textit{mutation}. The joint usage of these operators allows the algorithm to evolve a population of candidate solutions toward quasi-optimal solutions. A selection operator picks the ``best" individuals in the population, forming in this way the \textit{mating pool}. The individuals in the mating pool generate new possible solutions, forming the \textit{offspring}, through crossover and mutation operators, which respectively emulate the recombination and the mutation of individuals in a natural environment. In the end, the offspring replaces the initial population of the algorithm, and a new iteration (\textit{generation}) can start. The algorithm ends when a certain termination criterion is satisfied~\cite{yao:93}, for instance, when a maximum number of generations is reached or when no significant variation of the fitness is detected between successive iterations. 

A natural choice for the cost function is the total free energy~\cite{yang2020genetic}. However, this still requires evolving populations of entire twist and tilt distributions, parametrized in terms of the Cartesian components of the director at various locations between the substrates. This is essentially needed to approximate derivatives with finite differences, wherein the number of sampled points can severely affect the performance of the algorithm. Here, instead, we devise a GA to directly solve the bulk-integral version of the Euler-Lagrange equations~\cite{deuling:74,preist:89} in correspondence with a discrete set of positions within the cell, processed in a sequential routine. Remarkably, this approach allows us to model the response of the liquid-crystal layer to arbitrary field strengths.

In this paper, we focus on twisted nematic liquid-crystal (TNLC) cells, which feature complex three-dimensional modulations of the molecular director; however, the presented algorithm is not limited to TNLC configurations. First, we review the theory of TNLC cells in the presence of an external electric field, reporting the set of equations from which the twist and tilt distributions can be retrieved. Then, we provide technical details of our GA implementation and compute numerical results in a variety of settings. The solutions are also used to infer semi-analytical expressions that can provide good approximations for quantitative estimates. Finally, the performance of our routine is experimentally validated on spatially varying TNLC cells, specifically dual-$q$-plates~\cite{sit2023spatially}, fabricated by
designing different topological charges on the two alignment layers.

\section*{Theory}
The free energy density of the system, given by the fundamental elastic continuum equation, is~\cite{gennes:93},
    \begin{equation}
    \begin{split}
    \mathcal{F} 
        = \frac{1}{2} \{ &K_{1}[\div\mathbf{\hat{n}}(\mathbf{r})]^2 
       + K_{2}[\mathbf{\hat{n}}(\mathbf{r}) \cdot \curl \mathbf{\hat{n}}(\mathbf{r})]^2 \\
     &+ K_{3}[\mathbf{\hat{n}}(\mathbf{r}) \times \curl\mathbf{\hat{n}}(\mathbf{r})]^2
     -\mathbf{D}\cdot\mathbf{E} - \mathbf{B}\cdot\mathbf{H} \},
     \end{split}
    \end{equation}
where the $K_i$ are the elastic constants for the splay, twist, and bend distortions, respectively. The local director distribution $\mathbf{\hat{n}}(\mathbf{r})$ is our object of interest, and can be expressed in spherical coordinates,
    \begin{eqnarray}
        n_x &=& \cos{\phi(z)}\cos\theta(z), \nonumber \\
        n_y &=& \sin{\phi(z)}\cos\theta(z),  \\
        n_z &=& \sin\theta(z), \nonumber
    \end{eqnarray}
where $\phi(z)$ and $\theta(z)$ are the twist and tilt distributions, respectively, that minimize the free energy of the system. 
We limit the following discussion to the TNLC geometry. For the field-free case, one obtains a linear twist distribution ${\phi(z)=\alpha z/L}$, where $L$ is the thickness of the cell, and ${\theta(z)=0}$~\cite{yamauchi2005jones}. If an electric field is applied in the direction normal to the cell, $\mathbf{E}=E\mathbf{\hat{z}}$---which is equivalent to applying a voltage $V=EL$ across the twisted cell---the resulting $\phi(z)$ and $\theta(z)$ must be numerically found through a coupled set of integrals~\cite{deuling:74}. The twist distribution $\phi(z)$ is given by,
    \begin{equation}
        \phi(z) = \beta \int_0^{\theta(z)}\frac{\sqrt{1+\kappa\sin^2\theta}}{g(\theta)\cos^2\theta(1+\tau\sin^2\theta)}\text{d}\theta,
        \label{eqn:twisteqn}
    \end{equation}
where $\kappa=(K_3-K_1)/K_1$, $\tau=(K_3-K_2)/K_2$, and $\beta$ is an integration parameter to be determined from the boundary conditions. The tilt distribution $\theta(z)$ is determined implicitly through,
    \begin{equation}
        \frac{z}{L} = \frac{1}{2} \int_0^{\theta(z)}\frac{\sqrt{1+\kappa\sin^2\theta}}{g(\theta)} \text{d}\theta \bigg/ \int_0^{\theta_m} \frac{\sqrt{1+\kappa\sin^2\theta}}{g(\theta)}\text{d}\theta,
        \label{eqn:tilt}
    \end{equation}
where $\theta_m\equiv\theta_m(V)$ is the maximum tilt angle located at $z=L/2$ and is itself a function of the applied field. The $g(\theta)$ function depends on $\beta$ and $\theta_m$, with the form,
  \begin{equation}
        \begin{split}
            &g(\theta)=\bigg\{ \frac{\sin^2\theta_m-\sin^2\theta}{(1+\gamma\sin^2\theta)(1+\gamma\sin^2\theta_m)}+\beta^2 \frac{1+\kappa}{1+\tau}\times \\
            & \biggl(\frac{1}{(1+\tau\sin^2\theta_m)\cos^2\theta_m}-\frac{1}{(1+\tau\sin^2\theta)\cos^2\theta}\biggl)\bigg\}^{1/2},
        \end{split} \label{eq:g}
    \end{equation}
where $\gamma=(\epsilon_\parallel-\epsilon_\perp)/\epsilon_\perp$. Here, $\epsilon_\parallel$ ($\epsilon_\perp$) denotes the dielectric constant per unit volume that is parallel (perpendicular) to the local director. 
The maximum tilt angle $\theta_m$ is found from,
    \begin{equation}
        \frac{V}{V_{T0}} = \frac{2}{\pi}\int_0^{\theta_m} \frac{\sqrt{1+\kappa\sin^2\theta}}{(1+\gamma\sin^2\theta)g(\theta)}\text{d}\theta,
        \label{eqn:thetam}
    \end{equation}
where ${V_{T0}=\pi\sqrt{K_1/(\epsilon_0\Delta\epsilon)}}$, with ${\Delta\epsilon=\epsilon_\parallel-\epsilon_\perp}$, is the threshold voltage for ${\phi(z)=0}$, when the Fr\'eedericksz transition occurs in the zero-twist configuration~\cite{deuling:72}. The integration parameter $\beta$ can be found by evaluating Eq.~\eqref{eqn:twisteqn} at $z=L/2$:
    \begin{equation}
        \phi(L/2)=\frac{\phi_m}{2}=\beta\int_0^{\theta_m} \frac{\sqrt{1+\kappa\sin^2\theta}}{g(\theta)\cos^2\theta(1+\tau\sin^2\theta)}\text{d}\theta,
        \label{eqn:phim}
    \end{equation}
where $\phi_m$ is the maximum twist angle located at the back plate (${z=L}$) of the cell.

Equations~\eqref{eqn:thetam}-\eqref{eqn:phim} are a coupled set of integrals to be solved simultaneously. The numerical integration of this system is not trivial due to the strongly singular behaviour of the integrand functions. A better way to solve these equations is to do it iteratively, by first setting ${\beta=0}$ in Eqs.~\eqref{eq:g}-\eqref{eqn:thetam}, which produces a first estimate for ${\theta_{m0}=\theta_m(\beta=0)}$. This can be used to determine ${\beta_0=\beta(\theta_{m0})}$ from Eq.~\eqref{eqn:phim}, and so on until the desired convergence is achieved. Once the parameters $\beta$ and $\theta_m$ have been determined, we need to solve Eqs.~\eqref{eqn:twisteqn}-\eqref{eqn:tilt} to eventually extract the twist and tilt distributions. We have opted for an optimization approach based on evolutionary methods, specifically a genetic algorithm.

The following two analytical expressions~\cite{deuling:74} can be used as indicators for whether our GA converges successfully. The threshold voltage $V_T$ for a given total twist angle $\phi_m$ is,
    \begin{equation} \label{eq:vt}
        V_T(\phi_m) = V_{T0} \left[ 1 + \left( \frac{\phi_m}{\pi}\right)^2 \left( \frac{K_3}{K_1}-2\frac{K_2}{K_1}
 \right) \right]^{1/2},
    \end{equation}
and the value of the parameter $\beta_T$ at the threshold voltage for a given $\phi_m$ is,
    \begin{equation} \label{eq:beta-threhold}
        \beta_T(\phi_m) = \left[ \left( \frac{\pi}{\phi_m} \right)^2 +\frac{K_3}{K_1} - 2\frac{K_2}{K_1} \right]^{-1/2}.
    \end{equation}
These relations are useful as they describe limiting behaviours when ${V\rightarrow V_T}$, which our method must be able to reproduce. 

\section*{Numerical optimization}

To approximately solve Eqs.~\eqref{eqn:thetam}-\eqref{eqn:phim}, we implement two nested genetic algorithms which evolve real-valued individuals $\theta_{mi}$ and $\beta_i$. As prescribed by the iterative method mentioned above, the first GA is run to determine $\theta_{m0}$, corresponding to the initial guess for ${\beta=0}$. The second GA is then run to determine $\beta_0$, assuming ${\theta_m=\theta_{m0}}$, and so on. Each individual is a candidate to provide an optimal approximation to the actual solutions $\theta_m$ and $\beta$. By means of operators mimicking the natural selection mechanism, the GAs select for reproduction those individuals which better minimize the following cost functions:
\begin{equation}
        \mathcal{L}_{\theta_m} = \left| \frac{V}{V_{T0}} - \frac{2}{\pi}\int_0^{\theta_m} \frac{\sqrt{1+\kappa \sin^2\theta}}{(1+\gamma\sin^2\theta)g(\theta)}\text{d}\theta \right|^2, \label{eq:cost1}
\end{equation}
\begin{equation}
         \mathcal{L}_\beta = \left| \phi_m - 2\beta\int_0^{\theta_m} \frac{\sqrt{1+\kappa\sin^2\theta}}{g(\theta)\cos^2\theta(1+\tau\sin^2\theta)} \text{d}\theta  \right|^2, \label{eq:cost2}
\end{equation}
within the current generation. The numerical integrations required to evaluate these cost functions are successfully performed using the IMT-rule~\cite{iri:70} to handle singularities in finite integration regions~\cite{davis:14}. 

The detailed sequence of operators used in our algorithm will now be described. First, the well-known \textit{tournament selection} mechanism~\cite{goldberg:91} is used as a selection operator. This consists of repeating the following steps $N$ times, where $N$ is the population size:
    \begin{enumerate}
        \item Randomly select a subset of $k$ individuals.
        \item Choose the best individual in the subset to be inserted in the mating pool.
    \end{enumerate}
For our purposes, the best are those individuals that better minimize the cost functions in Eqs.~\eqref{eq:cost1}-\eqref{eq:cost2}. The \textit{blend crossover}~\cite{eshelman:93} is applied to mate individuals in the mating pool. When two individuals $\theta_A$ and $\theta_B$ reproduce, two newborn individuals $\theta_1$ and $\theta_2$ originate as random numbers belonging to the interval ${[\theta_A - c_i(\theta_B-\theta_A), \theta_B + c_i(\theta_B - \theta_A)]}$, where $c_i$ tunes the crossover, with ${i\in\{1,2\}}$, and we have assumed ${\theta_B\geq\theta_A}$. A similar reproduction occurs for two individuals $\beta_A$ and $\beta_B$. 

To explore a wider region of the parameter landscape, genetic mutations are included in the workflow in the form of Gaussian noise with mean $\mu$ and standard deviation $\sigma$, potentially affecting each newborn individual~\cite{kramer:17}. Our GAs also include an \textit{elitism} mechanism, i.e., the best individual from the old population is carried over to the next one, replacing the worst individual of the offspring. This pushes the algorithms to a faster convergence toward the best solutions. To preserve the physical validity of the final prediction for the maximum tilt angle $\theta_m$, a modulo-$\pi/2$ is performed after each operation on a $\theta$-individual. The maximum number of generations $N_{\text{gen}}$ is used as the termination criterion. Algorithm~\ref{alg:GA} presents the pseudo-code of the implemented GAs. Here, ${N=100}$, $N_{\text{gen}}=50$, ${k=4}$, ${c_1=c_2=0.5}$, ${\mu=0}$, and ${\sigma=0.2}$. Blend crossover and mutation are non-deterministic operators and occur with probability ${p_c=0.9}$ and ${p_m=0.01}$, respectively. The desired convergence for $\theta_m$ and $\beta$ is typically achieved within 10 iterations. Adequate convergence is considered to be when the cost functions are minimized with differences less than $10^{-15}$.

\begin{algorithm}[H] 
\caption{Pseudo-code of the implemented genetic algorithms}
\begin{algorithmic} 
\Require size of the population $pop\_size$, tournament size $k$, crossover probability $p_c$, $c$ for blend crossover, mutation probability $p_m$, $\mu$ and $\sigma$ for Gaussian mutation, termination criterion $t$
\Ensure the best solution $best$
\State $gen \gets 0$
\State $pop \gets $ generateRandomPopulation($pop\_size$)
\State checkPhysicalConstraints($pop$)
\State evaluateFitness($pop$)
\State $best \gets $ getBestIndividual($pop$)
\While{$gen < t$}
\State $offspring \gets$ executeTournament($pop,k$)
\State executeBlendCrossover($\textit{offspring}$, $p_c$, $c$)
\State checkPhysicalConstraints($offspring$)
\State executeGaussianMutation($offspring,p_m,\mu,\sigma$)
\State checkPhysicalConstraints($offspring$)
\State evaluateFitness($offspring$)
\State $pop \gets offspring$
\State $pop \gets$ elitism($pop,best$)
\State $best \gets$ getBestIndividual($pop$)
\State $gen \gets gen+1$
\EndWhile
\State \Return $best$
\end{algorithmic}
\label{alg:GA}
\end{algorithm}


Once the best estimates for $\theta_m$ and $\beta$ have been determined, a similar evolutionary strategy is devised for retrieving the tilt distribution, and then Eq.~\eqref{eqn:twisteqn} can be used to find the twist distribution.
The symmetry of the director around the mid-plane of the cell restricts the optimization to the first half of the cell: ${0<z<L/2}$. Half the cell thickness is divided into small intervals---here, 50 intervals are used---and the GA is executed within each slice. The set of solutions $\{ \theta(0), \theta(z_1),\theta(z_2),...,\theta(L/2) \}$ of Eq.~\eqref{eqn:tilt} is first determined by minimizing the following cost function within each interval:
    \begin{equation}
        \mathcal{L}_\theta = \left| \frac{z}{L}\int_0^{\theta_m}\frac{\sqrt{1+\kappa\sin^2\theta}}{g(\theta)}\text{d}\theta - \frac{1}{2}\int_0^{\theta(z)}\frac{\sqrt{1+\kappa\sin^2\theta}}{g(\theta)}\text{d}\theta \right|^2.
    \end{equation}
At each propagation distance $z_i$, the corresponding solution $\theta(z_i)$ is then used to determine $\phi(z_i)$ via Eq.~\eqref{eqn:twisteqn}.
Since the twist and tilt distributions are not expected to feature singular behaviours, the solution found at a given position $z_i$ cannot be too different from the solution associated with $z_{i-1}$ and $z_{i+1}$. Therefore, the initial population of the GA performed at each position other than $z=0$---which is known from boundary conditions---can be initialized from the solution found at the previous position, perturbed with a uniform noise $\Delta$. Here, $\Delta$ is chosen between 0.05 and 0.1, depending on how close to $V_T$ the current voltage is. This allows initializing the current GA very close to the actual solution. Accordingly, fewer generations are needed to obtain an adequate convergence, greatly reducing the computation time. The algorithms were performed using MATLAB R2021B on a laptop with an 11th Gen Intel\textregistered~Core\textsuperscript{TM} i5-1145G7 CPU @ 2.60~GHz, 2611~MHz, 4 cores, and 8 logical processors. The total calculation run time for a given set of initial conditions was about 2~min for the $\beta$--$\theta_m$ algorithm, and around 5~min for the $\theta$--$\phi$ algorithm.

\begin{figure}[t]
	\centering
	{\includegraphics[width=0.48\textwidth]{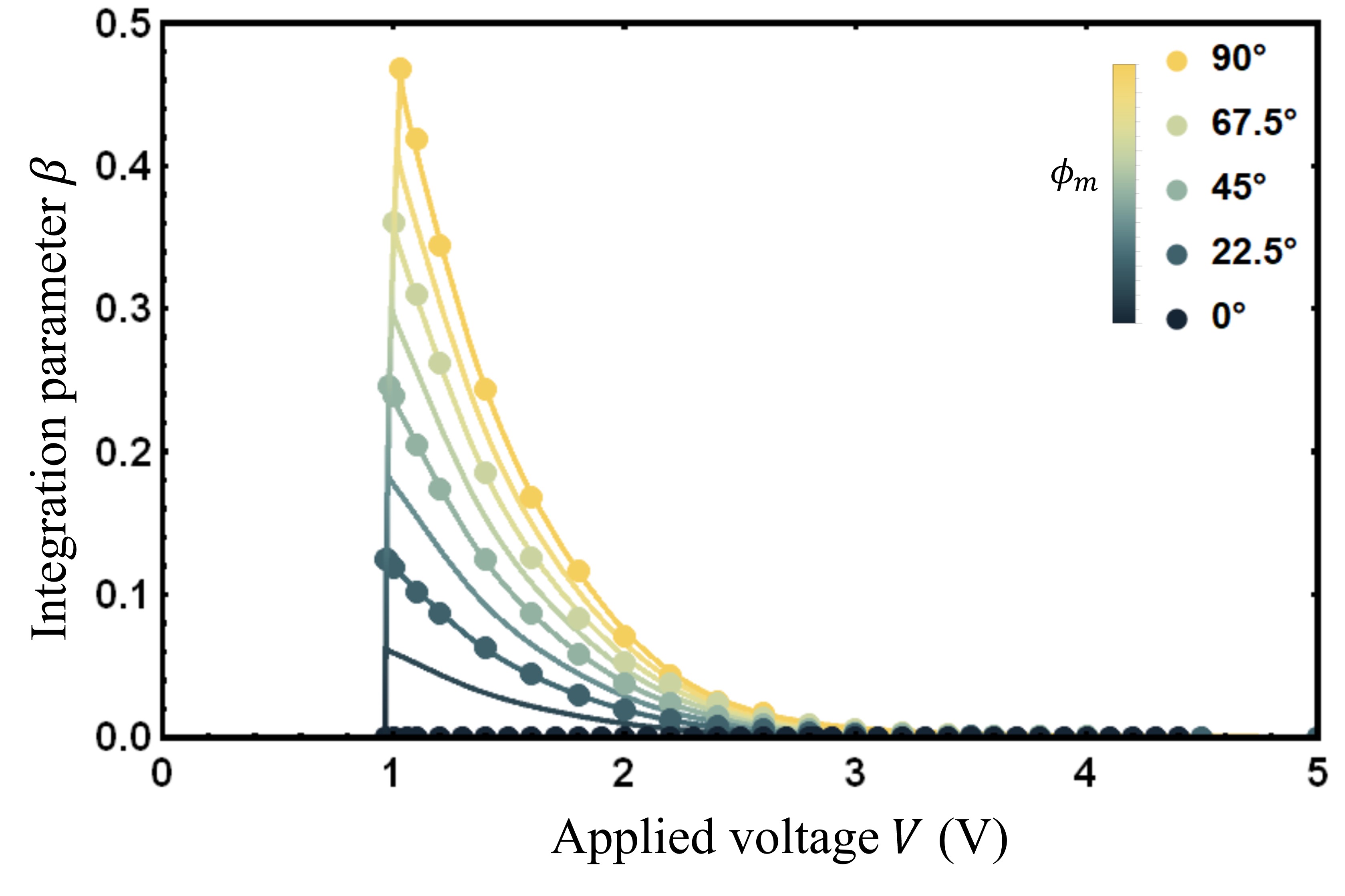}}
	\caption{\textbf{Integration parameter $\beta(\phi_m,V)$.} The dots correspond to the numerically calculated values for maximum twist angles ${\phi_m=90^\circ,67.5^\circ,45^\circ,22.5^\circ,0^\circ}$. The vertical gradient line is the analytical $\beta_T(\phi_m)$ values from Eq.~\eqref{eq:beta-threhold}, whereas the cascading solid-coloured lines are the lines of best fit using Eq.~\eqref{eq:beta-V}. }
	\label{fig:GA-beta-V}
\end{figure}

\begin{figure}[t]
	\centering
	{\includegraphics[width=0.48\textwidth]{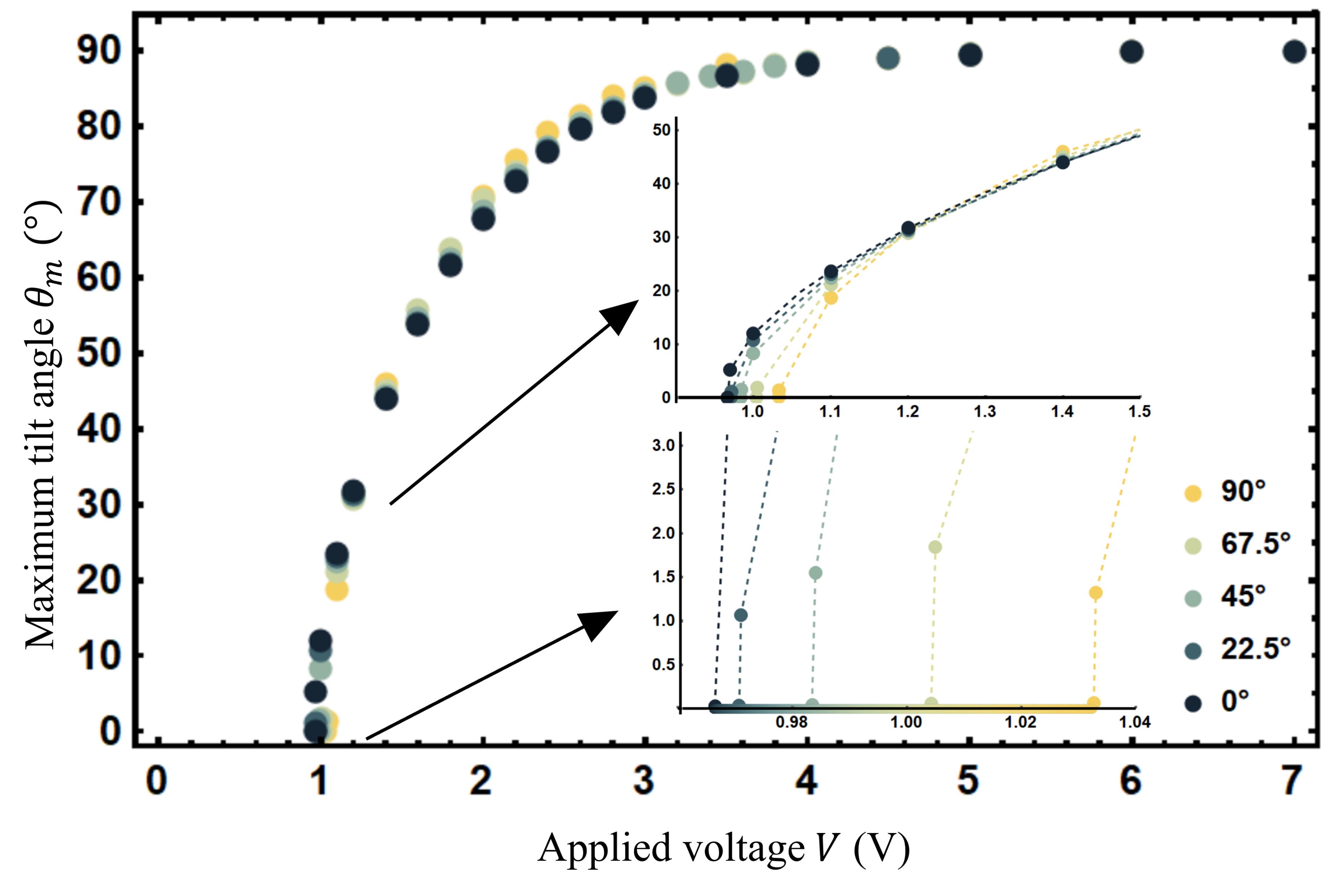}}
	\caption{\textbf{Maximum tilt angle $\theta_m(\phi_m,V)$.} The dots correspond to the numerically calculated values for maximum twist angles ${\phi_m=90^\circ,67.5^\circ,45^\circ,22.5^\circ,0^\circ}$. In the bottom inset, the gradient line is the analytical $V_T(\phi_m)$ values from Eq.~\eqref{eq:vt}. The dashed lines in both insets connect the dots for visual ease.}
	\label{fig:GA-theM-V}
\end{figure}

\section*{Numerical Results}
For our simulations, we use the material parameters for 6CHBT nematic liquid crystals, which have elastic constants of $K_1=6.7$~pN, $K_2=3.4$~pN, $K_3=10.6$~pN, and ${\Delta\epsilon=8}$~\cite{lc:12}. The zero-twist threshold voltage is $V_{T0}=0.966~V$. Figures~\ref{fig:GA-beta-V}-\ref{fig:GA-theM-V} report the numerically computed $\beta$ and $\theta_m$. The GA is run for maximum twist angles of ${\phi_m=90^\circ,~67.5^\circ,~45^\circ,22.5^\circ}$, and $0^\circ$, at a range of voltages from 0.96~V to 7~V, with resulting cost function $\mathcal{L}_\beta$ values of less than $10^{-32}$, and $\mathcal{L}_{\theta_m}$ between $10^{-17}$ and $10^{-26}$. 
For $\beta(\phi_m,V)$, the first measure of whether the GA is working is whether it can match the threshold $\beta_T(\phi_m)$ values given by Eq.~\eqref{eq:beta-threhold}. As shown in Fig.~\ref{fig:GA-beta-V}, the GA imitates the correct $\beta_T(\phi_m)$ at each maximum twist angle's threshold voltage. The data points for each $\phi_m$ follow a smooth decreasing trend, asymptotically approaching zero for large voltages. We propose a semi-analytical form to fit $\beta(\phi_m, V)$:
    \begin{equation} \label{eq:beta-V}
        \begin{split}
            \beta(\phi_m, V) = &\beta_T(\phi_m) \bigg( 1-\frac{2}{\pi} \times\\
            & \arctan\left[\sum_{i=1}^4 s_i V^i \sqrt{V-V_T(\phi_m)}\right] \bigg),
        \end{split}
    \end{equation}
where $s_i$ are a set of fitting parameters for all $\phi_m$.  Equation~\eqref{eq:beta-V} exhibits the expected behaviour: 

1. $\beta(\phi_m, V)$ does not exist for $V<V_T(\phi_m)$. 

2. When $V=V_T(\phi_m)$, then $ \beta(\phi_m, V)= \beta_T(\phi_m)$.

3. $\beta(\phi_m, V)\rightarrow 0$ as $V\rightarrow \infty$. 

The computed data set for $\beta(45^\circ,V)$ is used to obtain the fitting parameters $s_1=-7.95043$, $s_2=16.5784$, $s_3=-10.5245$, and $s_4=2.35869$, yielding a coefficient of determination ${R^2_{45} = 99.99\%}$.  
With these $s_i$, the coefficients of determination for the other data sets are ${R^2_{90}=99.98\%}$, ${R^2_{67.5}=99.97\%}$, ${R^2_{22.5}=99.98\%}$, and ${R^2_0 = 100\%}$. 
 
\begin{figure*}[t]
	\centering
	{\includegraphics[width=\textwidth]{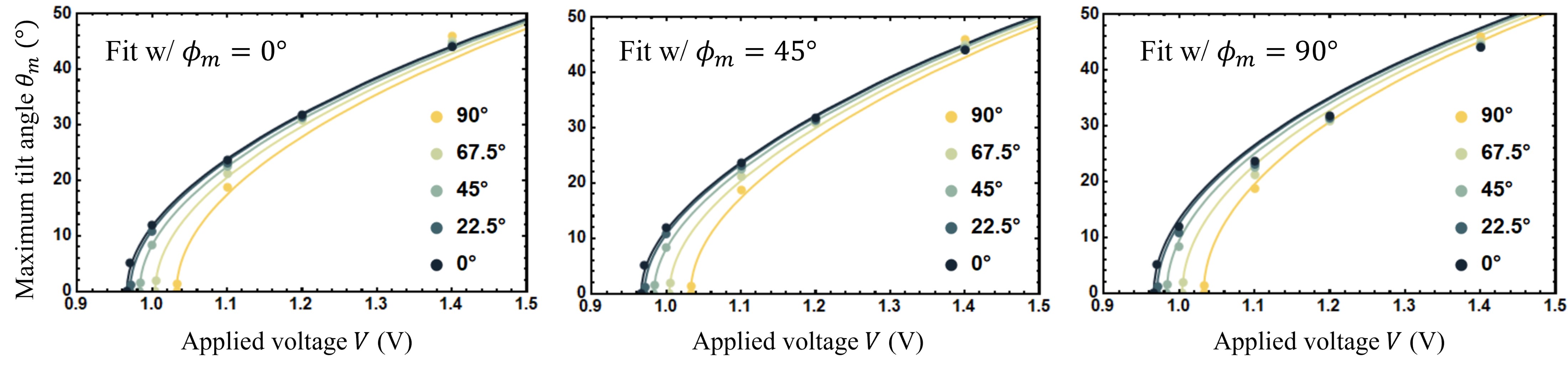}}
	\caption{\textbf{Fits for $\theta_m(\phi_m,V)$.} With the ansatz of Eq.~\eqref{eq:theM-ansatz-1}, fits are produced using the ${\phi_m=0^\circ,45^\circ,90^\circ}$ data sets. Each fit is then used to produce the $\theta_m(\phi_m,V)$ curves for other $\phi_m$ values, as shown in each subplot.}
	\label{fig:theM-fits}
\end{figure*} 

\begin{figure*}[t]
	\centering
	{\includegraphics[width=0.99\textwidth]{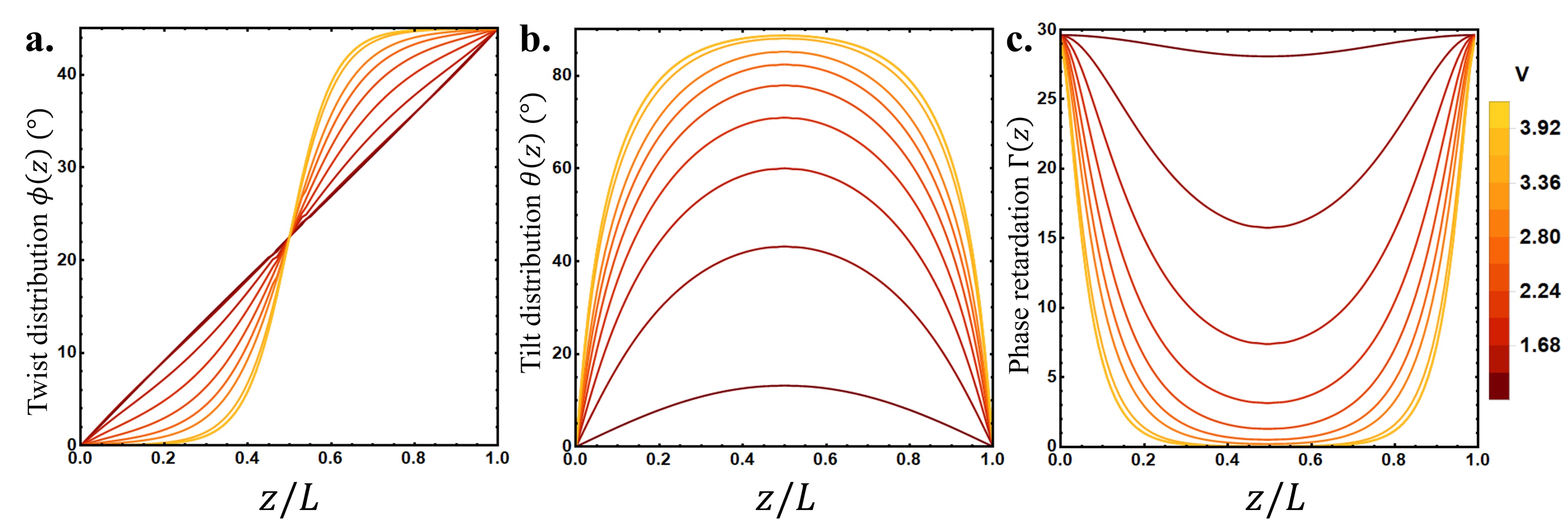}}
	\caption{\textbf{GA results for ${\phi_m=45^\circ}$.} \textbf{a}.~Twist $\phi(z)$, \textbf{b}.~tilt $\theta(z)$, and \textbf{c}.~phase retardation $\Gamma(z)$ distributions numerically calculated for a range of voltages between 1.0~V and 4.2~V. The $\Gamma(z)$ are computed using $L=35~\mu$m, $\Delta n=0.151$, and $\lambda = 632$~nm.}
	\label{fig:GA-pi-4-diffV}
\end{figure*} 

For the maximum tilt angle $\theta_m(\phi_m,V)$, we expect ${\theta_m=0}$ when ${V<V_T}$ for each $\phi_m$. The bottom inset of Fig.~\ref{fig:GA-theM-V} shows that our GA reproduces the threshold voltage values of Eq.~\eqref{eq:vt}. The top inset of Fig.~\ref{fig:GA-theM-V} is a zoom-in to show what appears to be a common crossing point around 1.2~V with $\theta_m \sim 30.9^\circ$. These curves can each be fit with the form,
    \begin{equation} \label{eq:theM-ansatz-1}
        \theta_m(\phi_m,V) = \arctan\left[ \sum_{i=1}^4 b_iV^i \sqrt{V-V_T(\phi_m)}   \right],
    \end{equation}
where $b_i$ are a new set of fitting parameters. This ansatz obeys the expected behaviour: 

1. When $V=V_T(\phi_m)$, then $\theta_m(\phi_m,V)=0$.

2. $\theta_m(\phi_m,V) \rightarrow 90^\circ$ as $V\rightarrow \infty$. 

However, this form fails to reproduce the crossing point around 1.2~V  for the fitted data set, despite the corresponding $R^2$ values being 99.99\%, as can be seen in Fig.~\ref{fig:theM-fits}. 
This suggests that there is an extra $\phi_m$-dependence not included in Eq.~\eqref{eq:theM-ansatz-1}. 
A hint about this dependence can be found in Ref.~\cite{preist:89}, which studies the twist and tilt distributions in the high-voltage limit ${V\gg V_{T0}}$:
    \begin{equation}
        \tan^2(\theta_m) \approx \left( 1+ \tan^2(\phi_m/2)  \right) \tan^2(\theta_{m}^{(0)}),
    \end{equation}
where $\theta_{m}^{(0)}$ is the maximum tilt angle for a non-twisted cell at the same voltage. However, an analytical expression for voltages below or near the threshold was not found in the literature. If a semi-analytical form could be found that reproduces the GA outputs for $\theta_m(\phi_m,V)$, then along with Eq.~\eqref{eq:beta-V} for $\beta(\phi_m,V)$ one would no longer need to run the GA for every $(\phi_m,V)$ configuration, dramatically reducing computation time.

\begin{figure*}[]
	\centering
	{\includegraphics[width=\textwidth]{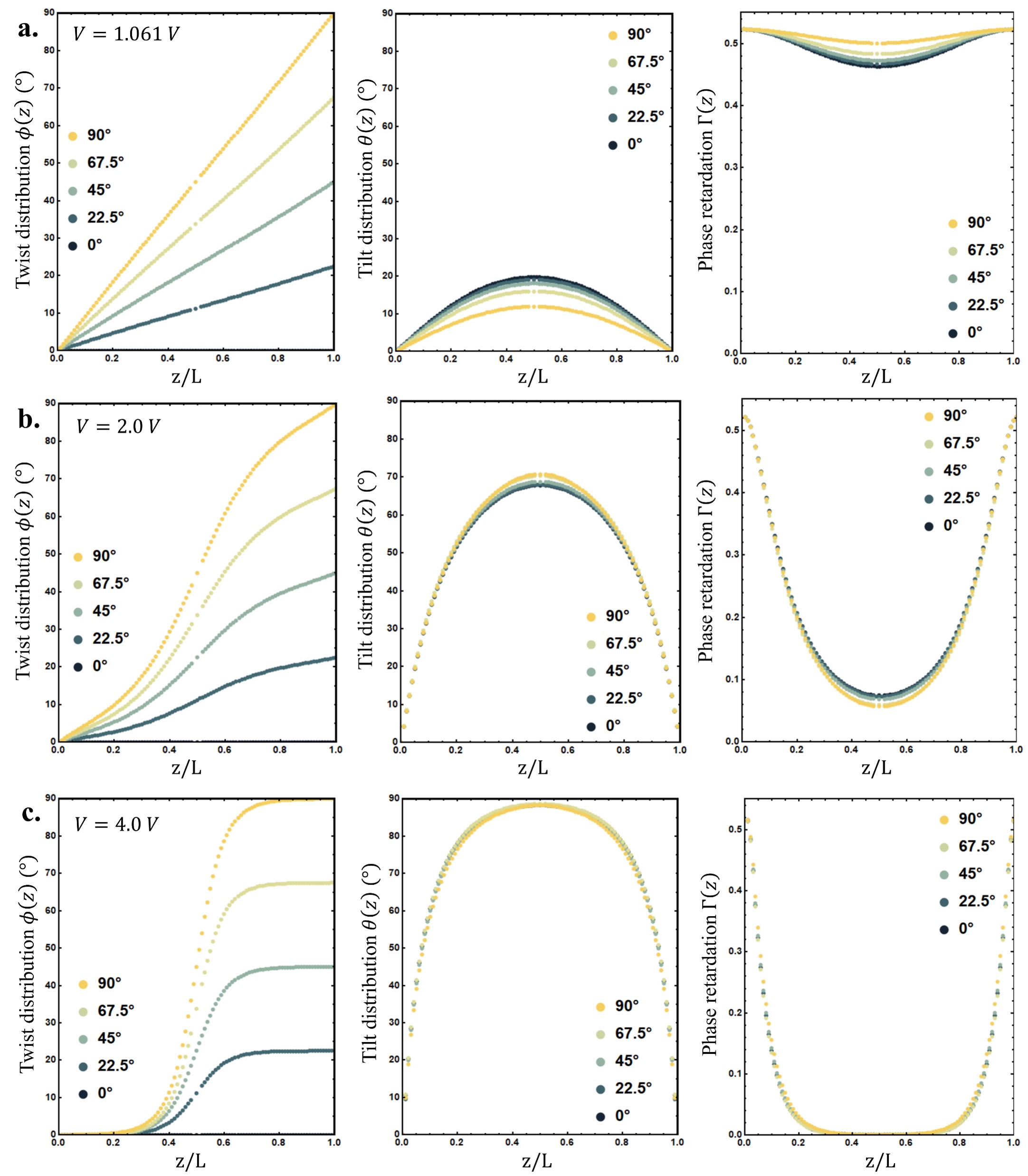}}
	\caption{\textbf{GA results for different maximum twist angles.}  Twist, tilt, and phase retardation distributions at \textbf{a}.~$V=1.061$~V, \textbf{b}.~$V=2.0$~V, \textbf{c}.~$V=4.0$~V.}
	\label{fig:GA-diffPhimV}
\end{figure*}

Figure~\ref{fig:GA-pi-4-diffV}\textbf{a}-\textbf{b} reports an example of the numerically obtained twist and tilt distributions for ${\phi_m=45^\circ}$ with various voltage settings. The twist distribution is no longer linear as we increase the applied field strength (see Fig.~\ref{fig:GA-pi-4-diffV}\textbf{a}), as was first calculated by Deuling~\cite{deuling:74}. Nevertheless, there appears to be a voltage above the threshold up to which $\phi(z)$ is essentially still linear. As the field strength increases further, the distributions feature an S-like shape, becoming sharper and more step-like. As expected, the liquid crystals start tilting in the direction of the applied field only above the threshold voltage (see Fig.~\ref{fig:GA-pi-4-diffV}\textbf{b})~\cite{deuling:74}.
Above the threshold, the maximum tilt angle steadily approaches $\theta_m=90^\circ$ with increasing voltage. For very intense fields, the distribution flattens out in the middle. The phase retardation $\Gamma(z_j)$ within the $j^{\text{th}}$ slice in Fig.~\ref{fig:GA-pi-4-diffV}\textbf{c} is calculated from the tilt distribution using the trapezoidal rule in favour of left or right Riemann sums for a better estimate,
    \begin{equation} \label{eq:gammaTrap}
       \Gamma(z_{j}) = \frac{\pi\Delta n d}{\lambda}[\cos^2\theta(z_{j+1})+\cos^2\theta(z_j) ].
   \end{equation}
A cell thickness of $L=35~\mu$m is used, with ${N=100}$ slices for ${d=L/N}$, $\Delta n = 0.151$, and $\lambda = 632$~nm. The twist and tilt distributions, along with the phase retardation, have also been extracted for different maximum twist angles $\phi_m$ at a range of voltages (see Fig.~\ref{fig:GA-diffPhimV}). 

\begin{figure*}[t]
	\centering
	{\includegraphics[width=0.85\textwidth]{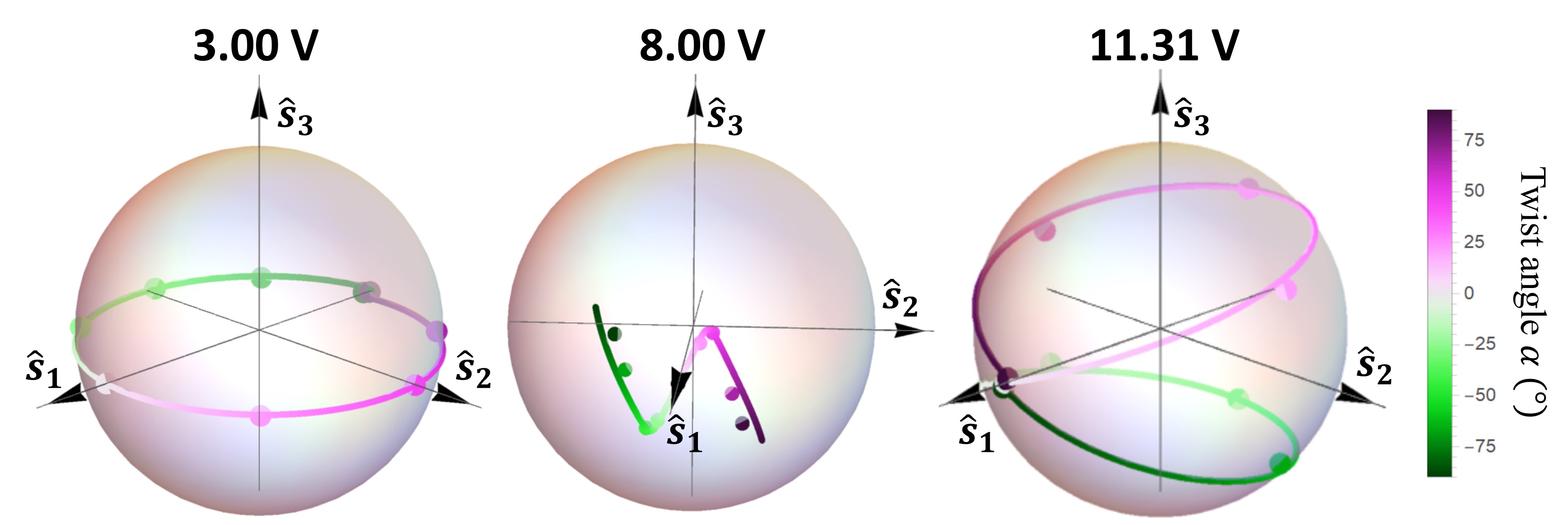}}
	\caption{\textbf{Validation of the twist scaling approximation}. Comparison of output states generated via the approximated twist and tilt distributions (colored lines), based on ${\phi(45^\circ,V,z)}$ and $\theta(45^\circ,V,z)$, and the numerically calculated distributions for each $\phi_m$ (colored dots), at applied voltages of $V=3.00$, 8.00, and 11.31$~V$. A horizontally polarized input was considered.}
	\label{fig:GA-distr-comp}
\end{figure*} 

Using the numerically calculated $\phi(z)$, $\theta(z)$, and Eq.~\eqref{eq:gammaTrap}, we can derive the total Jones matrix of a TNLC cell for any given $\phi_m$ and applied voltage $V$. In the case with no external field, the device is modelled as a stack of $N$ linearly twisted cells of thickness $d$, with constant phase retardation. 
The Jones matrix $\mathbf{T}_{\phi_f}(\phi_m,\Gamma)$ for a given ${\phi_m = \phi_b -\phi_f}$ and phase retardation $\Gamma$ is given by~\cite{yariv:84},
    \begin{eqnarray}
        \mathbf{T}_{\phi_f}(\phi_m,\Gamma)
        &=& \mathbf{R}(-\phi_b)\mathbf{M}_0(\phi_m,\Gamma)\mathbf{R}(\phi_f), \label{eq:T1}
    \end{eqnarray}
where, 
    \begin{equation}
        \mathbf{M}_0(\phi_m,\Gamma)= \begin{bmatrix}
            \cos X -\frac{i\Gamma}{2X}\sin X & \frac{\phi_m}{X}\sin X   \\
             -\frac{\phi_m}{X}\sin X   & \cos X +\frac{i\Gamma}{2X}\sin X 
        \end{bmatrix},
    \end{equation}
$X=\sqrt{\phi_m^2 +(\Gamma/2)^2}$, $\phi_f=\phi(0)$, and $\phi_b=\phi(L)$ are the front and back alignment angles, respectively, and $\mathbf{R}(\cdot)$ is the rotation matrix,
    \begin{equation}
        \mathbf{R}(\cdot) = \begin{bmatrix}
                   \cos(\cdot) & \sin(\cdot) \\
                   -\sin(\cdot) & \cos(\cdot)
                \end{bmatrix}.
    \end{equation} 
When a voltage $V$ is applied across the cell, the total Jones matrix can be then approximated as,
    \begin{eqnarray} \label{eq:jm-Num}
        \mathbf{J}_{\phi_f}(\phi_m,V) &=& \prod_{j=0}^N \mathbf{T}_{\phi(z_j)} \left( \phi_m(z_j), \Gamma(z_j) \right) \\
        &=& \mathbf{R}(-\phi_b) \left[ \prod_{j=0}^N \mathbf{M}_0(\phi_m(z_j),\Gamma(z_j)) \right] \mathbf{R}(\phi_f), \nonumber
    \end{eqnarray}
where $z_j=jd$, $\phi_m(z_j)=\phi(z_{j+1})-\phi(z_{j})$.

To further save on computation time, we make the additional approximation that the twist distributions for different $\phi_m$ are scaled versions of each other, e.g., $\phi(\phi_m,V,z)=(\phi_m/45^\circ)\times\phi(45^\circ,V,z)$. From the various twist distributions shown in the left column of Fig.~\ref{fig:GA-diffPhimV}, this appears to be a reasonable assumption. Consequently, the tilt and phase retardation distributions are assumed to be the same for all $\phi_m$ at a given voltage, i.e., ${\theta(\phi_m,V,z) = \theta(45^\circ,V,z)}$, and ${\Gamma(\phi_m,V,Z) = \Gamma(45^\circ,V,z)}$. This is particularly the case for high voltages, with deviations expected at lower voltages (see Fig.~\ref{fig:GA-diffPhimV}). Figure~\ref{fig:GA-distr-comp} compares the output polarizations plotted on the Poincar\'e Sphere at different voltages for the full numerically calculated distributions for ${\phi_m=-90^\circ}$, $-67.5^\circ$, $-45^\circ$, $-22.5^\circ$, $0^\circ$, $22.5^\circ$, $45^\circ$, $67.5^\circ$, $90^\circ$, with the approximated distributions based on ${\phi_m=45^\circ}$. The simulations are obtained from Eq.~\eqref{eq:jm-Num}, assuming a horizontally polarized input state. The differences are minimal, with average state overlaps ${(1+\mathbf{S}_{\phi_m}\cdot\tilde{\mathbf{S}}_{\phi_m})/2}$ of over $99\%$, where $\mathbf{S}_{\phi_m}$ and $\tilde{\mathbf{S}}_{\phi_m}$ are the output Stokes vector resulting from the numerically calculated and approximated distributions, respectively, for a given $\phi_m$.

\begin{figure}[t]
	\centering
	{\includegraphics[width=0.45\textwidth]{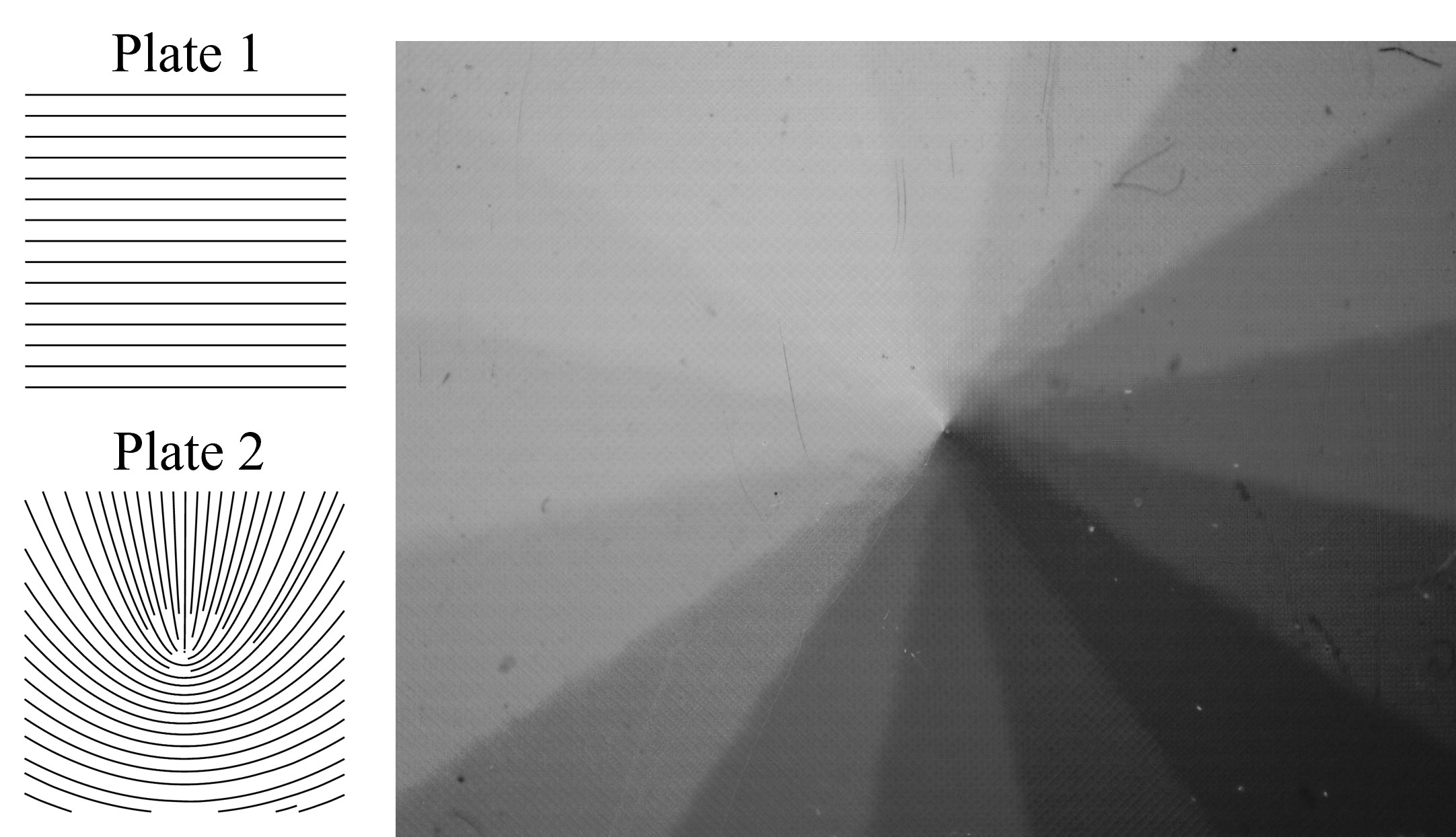}}
	\caption{\textbf{Fabricated sample.} Image of DP(0,1/2) between crossed polarizers under a microscope, illuminated with white light. The topological pattern on each glass plate is also shown. The ${q=1/2}$ pattern is discretized into 16 slices to explore maximum twist angles $\phi_m$ from $-90^\circ$ to $90^\circ$.}
	\label{fig:dp}
\end{figure} 

\begin{figure*}[]
	\centering
	{\includegraphics[width=\textwidth]{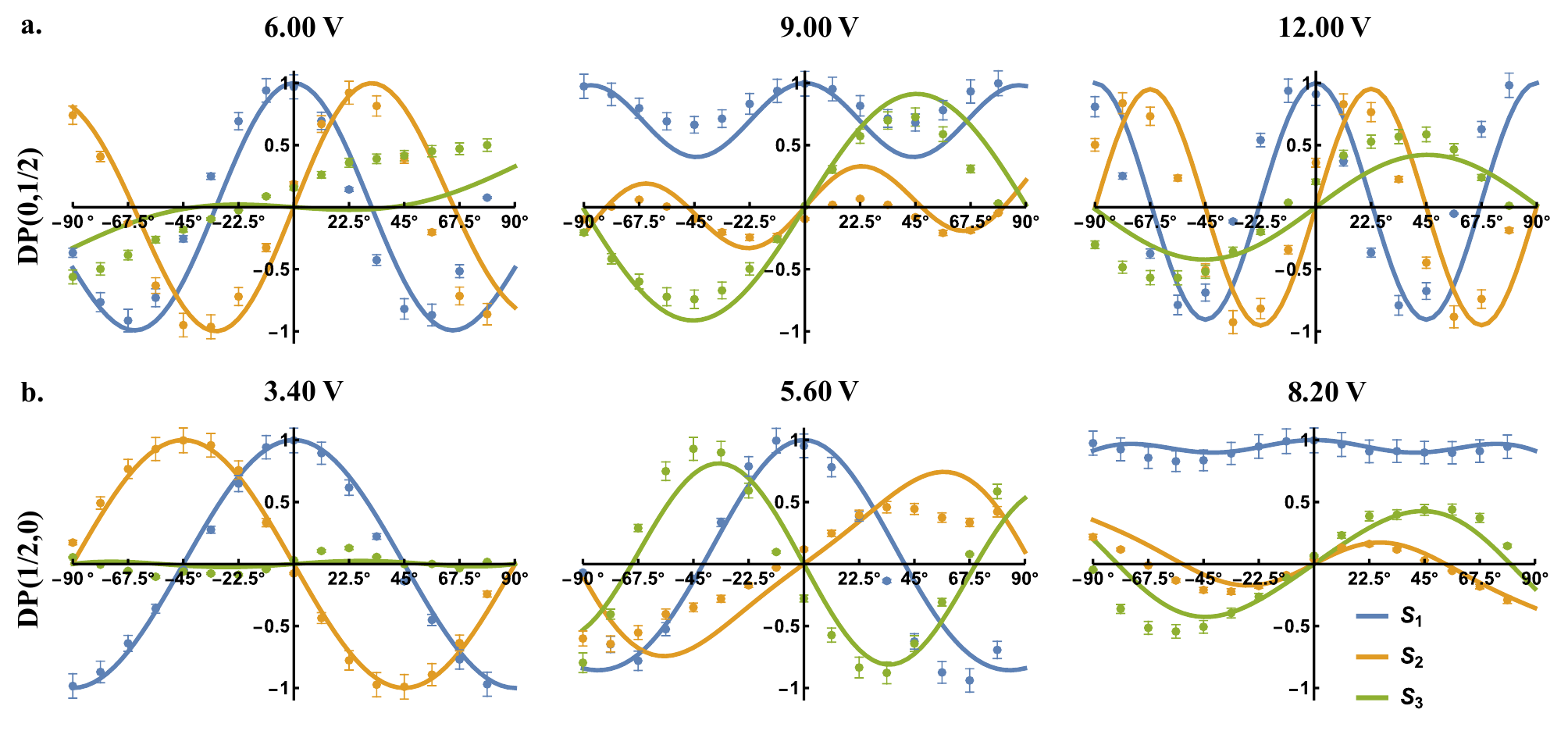}}
	\caption{\textbf{Experimental results.} Comparison between numerical predictions and experimentally reconstructed Stokes parameters for \textbf{a}.~DP(0,1/2), at $V_{pp}=6.00$, 9.00, and 12.00~$V$, and \textbf{b}.~DP(0,1/2), at $V_{pp}=3.40$, 5.60, and 8.30~$V$.}
	\label{fig:comp}
\end{figure*}

\section*{Experimental Results}
The predictions of our numerical routine are tested with TNLC plates with a spatially varying maximum twist angle (see Ref.~\cite{sit2023spatially} for details). In the following, we briefly review the fabrication technique of these devices. Two glass plates with a thin layer of conductive indium tin oxide (ITO) are coated with an azobenzene-based dye. The dye molecules are photoaligned when exposed to linearly polarized light at a wavelength within the peak of the absorption spectrum. In our realization, we separately patterned the glass plates such that the front plate features a uniform alignment, and the back plate is patterned with a ${q=1/2}$ topology which has been discretized into 16 slices. Finally, the two plates are sealed and the nematic liquid crystals (6CHBT) penetrate into the sample by capillarity, exhibiting a full range of maximum twist angles $\phi_m$ from $-90^\circ$ to $90^\circ$. Figure~\ref{fig:dp} shows the fabricated sample between crossed polarizers under a microscope, illuminated with white light. We refer to these inhomogeneous non-symmetrically patterned devices as dual-plates (DPs), as they exhibit a different behaviour depending on the plate orientation. For the DP used here, the configuration in which light passes through the uniform pattern first and exits through the $q=1/2$ pattern is denoted as DP(0,1/2), and \textit{vice versa} as DP(1/2,0). 

The action of each configuration on polarized light is characterized via polarization tomography, to reconstruct the output polarization distribution when different voltages are applied across the cell. Figure~\ref{fig:comp}\textbf{a}-\textbf{b} compares the experimentally reconstructed Stokes parameters with the predicted outputs obtained from the approximated total Jones matrices of DP(0,1/2) and DP(1/2,0), respectively, at different voltages. In our experiment, a sinusoidal waveform with 4~kHz frequency was used. An excellent agreement is observed in all realizations, with average overlaps of $93\pm2\%$, $98.0\pm0.3\%$, and $92.7\pm0.4\%$ for DP(0,1/2), at voltages ${V_{pp}=6.00}$, $9.00$ and $12.00$~V, respectively (see Fig.~\ref{fig:comp}\textbf{a}), where $V_{pp}$ is the peak-to-peak voltage, and the average is computed over the outputs within each of the 16 slices. For DP(1/2,0), we obtain $99.7\pm0.1\%$, $95\pm1\%$, and $99.2\pm0.3\%$, at ${V_{pp}=3.40}$, $5.60$ and $8.20$~V, respectively (see Fig.~\ref{fig:comp}\textbf{b}). Deviations from numerical predictions are mainly ascribed to fabrication defects and environmental temperature fluctuations, which can slightly change the liquid-crystal intrinsic birefringence and elastic constants. These results certify that our numerical routines are suitable for predicting the optical action of individual devices with high accuracy.

\section*{Conclusion}
We have demonstrated a novel robust approach to the determination of twist and tilt distributions of liquid-crystal cells in the presence of an external field. Our method directly tackles the integral form of the Euler-Lagrange equations, thereby avoiding the necessity of trial solutions. The complexity of the equation system is fragmented within subsequent genetic routines, each of which uses the outputs of the previous one to converge to the optimal solutions. Our method has been validated both numerically and experimentally on two configurations of spatially varying TNLC cells, where the optimization runs over multiple transverse positions. This scheme can provide a useful tool for the experimental characterization of the next generations of dual-devices, such as dual-lenses and gratings. At the same time, it will be interesting to explore machine-learning approaches to extract the liquid-crystal director distributions in real-time~\cite{sigaki2020ML,piven2024ML}.

\section*{Acknowledgements}
A.S. acknowledges the financial support of the Vanier graduate scholarship of the NSERC. This work was supported by the Ontario's Early Researcher Award (ERA), Canada Research Chairs (CRC) and Natural Sciences and Engineering Research Council of Canada (NSERC). 

\bibliography{uo-ethesis}

\end{document}